\documentclass{INTERSPEECH2023}

\interspeechcameraready 

\usepackage{multirow}

\title{
Self-Supervised Acoustic Word Embedding Learning via Correspondence Transformer Encoder
}
\name{Jingru Lin$^1$, Xianghu Yue$^1$, Junyi Ao$^2$, Haizhou Li$^{1,2}$}
\address{
  $^1$Department of Electrical and Computer Engineering,
National University of Singapore, Singapore\\
  $^2$Shenzhen Research Institute of Big Data, School of Data Science, The Chinese University of Hong Kong, Shenzhen, China}
\email{ \{jingrulin, xianghu.yue\}@u.nus.edu, junyiao1@link.cuhk.edu.cn, haizhouli@cuhk.edu.cn}

\begin{document}

\maketitle
 
\begin{abstract}
Acoustic word embeddings (AWEs) aims to map a variable-length speech segment into a fixed-dimensional representation. High-quality AWEs should be invariant to variations, such as duration, pitch and speaker. In this paper, we introduce a novel self-supervised method to learn robust AWEs from a large-scale unlabelled speech corpus. Our model, named Correspondence Transformer Encoder (CTE), employs a teacher-student learning framework. We train the model based on the idea that different realisations of the same word should be close in the underlying embedding space. Specifically, we feed the teacher and student encoder with different acoustic instances of the same word and pre-train the model with a word-level loss. Our experiments show that the embeddings extracted from the proposed CTE model are robust to speech variations, e.g. speakers and domains. Additionally, when evaluated on Xitsonga, a low-resource cross-lingual setting, the CTE model achieves new state-of-the-art performance.

\end{abstract}

\vspace{-0.2cm}
\section{Introduction}
Mapping arbitrary-length words into fixed-dimensional vector representations is very useful for many speech processing tasks~\cite{chung2016audio}, such as query-by-example~\cite{settle2017query,yuan2018learning} and speech recognition~\cite{settle2019acoustically,bengio2014word}. 
Once word segments are represented as fixed-dimensional vectors, they can be compared through simple cosine or Euclidean distance efficiently, or directly applied to downstream tasks' classifiers~\cite{yang2022KF,ye2023partial,yang2022deep}.

The earliest works try to embed variable-length acoustic words into fixed-dimensional vectors using simple heuristic approaches such as down-sampling~\cite{levin2013fixed}.
However, these vector representations may not be able to precisely describe the structure of the audio segments for the reason that the speech instances of the same word will never be identical due to the variations in duration, pitch, speakers' accent and gender, etc.
This makes learning acoustic word embeddings (AWEs) more challenging than textual word  embeddings~\cite{mikolov2013distributed,mikolov2013efficient}.
The former should generate similar representations from different realisations of the same acoustic word despite the variations, while the latter only needs to describe one unique character sequence for each word.

On the other hand, deep learning, which seeks to learn from the data, has been more successful in describing the acoustic word structures against the variations~\cite{gao2023self}. Some earlier works that applied deep learning to learn AWEs relied on using the true word identity~\cite{maas2012word, bengio2014word, chen2015query}. 
Nonetheless, learning from a large amount of annotated data is not only expensive but also contradictory to the way that human infants first acquire languages where little supervision is needed~\cite{chen2019audio}.
This drives the research for AWEs towards unsupervised learning where AWEs are learned from unannotated speech data. In unsupervised settings, the boundaries of words in a speech sentence and the identity of the acoustic words are unknown. Therefore, the study of AWEs in unsupervised settings has practical implications for zero- or low-resource settings where transcribed data is unavailable or very scarce. 

Generally, in the unsupervised setting, only word pairs ('same' or 'different' to indicate if a word pair has the same or different identity) or word boundary information are needed as a form of weak supervision. 
For example, Kamper~\textit{et al.} applies convolution neural networks in a Siamese setting~\cite{kamper2016deep}, where the Siamese networks learn to maximise/minimise the distance between words of different/same types; other works~\cite{kamper2015unsupervised,kamper2019truly} also use word pairs information but with different model architectures.
Chung~\textit{et al.} segments speech sentences into acoustic words based on word boundary information and implements an autoencoder model, which encodes the extracted acoustic words input into fixed-length representations, and then reconstructs this word input out from the representations~\cite{chung2016audio}.
Although these works make use of weak supervision during training, advances in zero-resource technology makes it possible to obtain the word pairs and boundary information from speech corpus in unsupervised ways~\cite{jansen2011efficient,rasanen2015unsupervised}. To further improve the acoustic embeddings, some researchers exploit the use of the words' character sequences~\cite{he2016multi,jung2019additional}.
However, the use of character sequences violates the purpose of unsupervised learning and hinder the applications to the low- or zero-resource setting as the transcribed data is not easily available in these settings.

In this work, we extend the research on unsupervised AWEs. 
In contrast to the above approaches where complex models are trained in attempts to model as much information from the limited target resources as possible, here we want to learn a general representation of acoustic words from a larger resource. 
When applied to low- or zero-resource settings, the models should work as a robust feature extractor and efficiently adapt to new resources.
For this purpose, we propose Correspondence Transformer Encoder (CTE), a novel self-supervised technique to learn general and robust AWEs by leveraging large amounts of unlabelled speech data. Our CTE is built based on the teacher-student learning framework.
Specifically, CTE has a student and a teacher encoder that share the same architecture.
During pre-training, the student and teacher encoder each takes an acoustic instance of the same word. 
The student encoder is optimized to minimise the distance between the output representations generated by itself and the teacher encoder, and the teacher encoder is parameterised by an exponentially moving average (EMA) of the student network. 
In downstream evaluations, the representations generated by the student encoder are used as word embeddings. 

We verify the effectiveness of our model on both in-, cross-domain, and cross-lingual settings. Experimental results show that our proposed CTE learns more robust word representations while discarding other irrelevant information that is sensitive to variations in speech. To the best of our knowledge, this is the first work that utilises a transformer architecture to learn acoustic word embeddings.


\vspace{-0.1cm}
\section{Methodology}
\label{sec:approach}

\begin{figure}[t]
 
\begin{minipage}[t]{1.0\linewidth}
  \centering
  \centerline{\includegraphics[width=7.5cm]{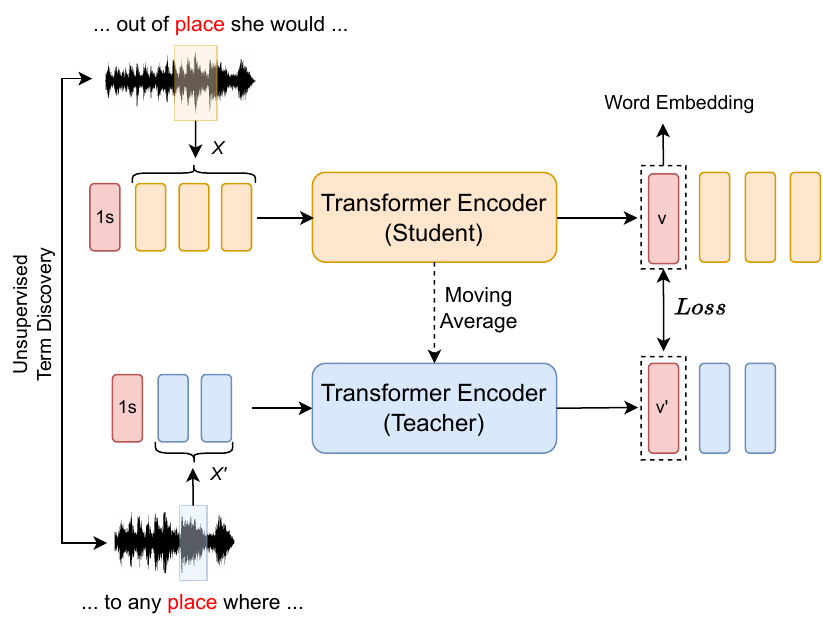}}
  \caption{The overall model architecture of CTE. The teacher encoder and student encoder are fed with different acoustic instances of the same word. The teacher is parameterised by an exponentially moving average of the student weights, while the student predicts the average representation of top K layers from the teacher.}
  \label{fig:model}
  \vspace{-0.2cm}
\end{minipage}
\end{figure}

The goal of our AWE models is to learn a function $f$ such that $f(x)$ can map variable-length acoustic word segment $x$ into a fixed-dimensional word embedding $v$. It is desired that different acoustic segments representing the same word should be mapped to close proximity in the embeddings space, regardless of their variations. Below we formally present the CTE model.

\subsection{Model architecture}
\label{ssec:model}
The overall architecture of the proposed correspondence transformer encoder CTE is shown in Fig.~\ref{fig:model}. Our motivation comes from the premise that different acoustic realisations of the same word should be close in the underlying embedding space.

For CTE, a single training input consists of a pair of audio segments that have the same word identity. 
Given the pair of audio segments, we first extract the 80-dimensional log Mel-filterbank, denoted as $(X, X')$, where $X=[x_1, \dots, x_{t_1}]$ and $X'=[x'_1, \dots, x'_{t_2}]$ with lengths $t_1$ and $t_2$, respectively.
We pass the acoustic features $X$ to the student encoder and $X'$ to the teacher encoder.
The student and teacher encoders are both transformer encoders, and the teacher parameters are an exponentially moving average of the student weights.

To obtain fixed-dimensional word-level representations, taking inspiration from BERT, we add a random vector at the first timestep of the input features and use the first-timestep vector from the representation as the final word embeddings. In practise, the first-timestep random vector can be initialised with all ones or zeros. Here, we only use ones to denote. The representations obtained from $X$ and $X'$ are hence given as:
\begin{equation}
    [v, h] = Encoder_{student}([\mathbf{1}, X]) 
\end{equation}

\vspace{-0.3cm}
\begin{equation}
    [v', h'] = Encoder_{teacher}([\mathbf{1}, X'])
\end{equation}
where $Encoder_{student}$ and $Encoder_{teacher}$ are the student and teacher encoder respectively, $\mathbf{1}$ is the all-ones vector added to the first timestep of the input features, $v$ and $v'$ are the first-timestep vectors  of the encoded representations which are regarded as the respective acoustic embeddings of the input word pair $X$ and $X'$.  

\subsection{Self-supervised learning task}
\label{ssec:ssl}
Given $(X, X')$, the model is trained to learn the word embedding $v$ from $X$ that can predict the word embedding $v'$ from $X'$. This self-supervised learning target $v'$ is constructed based on the first timestep of the output representations from the top $K$ blocks of the teacher encoder. We first apply a layer normalization to each block, and then average the top $K$ blocks to construct $v'$:
\vspace{-0.2cm}
\begin{equation}
    v' = \frac{1}{K} \sum_{l=L-K+1}^{L} v_l
\end{equation}
\vspace{-0.1cm}
where $v_l$ is the first-timestep vector from $l$-th layer in the teacher encoder and $L$ is the total number of layers in the encoder. The vector $v'$ is to be predicted by the student encoder. Hence, CTE is optimized by minimising the cosine distance between the vector representations $v$ and $v'$. The training loss $\mathcal{L}$ is given by:
\vspace{-0.1cm}
\begin{equation}
    \mathcal{L}=1-cos(v, v')=1-\frac{v \cdot v'}{||v||_2 \cdot ||v'||_2}
\end{equation}
\vspace{-0.1cm}

In CTE, the student and teacher encoders share the same architecture, but differ in the way they are parameterised. The training loss $\mathcal{L}$ updates the parameters $\theta$ of the student encoder while the teacher encoder is parameterised by an exponentially moving average (EMA) of the student parameters. This means, given a target decay rate $\tau$, after each training step, the teacher encoder parameters $\xi$ are given by:
\vspace{-0.2cm}
\begin{equation}
    \xi = \tau \xi + (1 - \tau)\theta
\end{equation}

\subsection{Word embeddings}
In the inference stage, the teacher encoder is discarded and only the student encoder is needed to produce the word embeddings. Given an audio segment, the student encoder takes in its 80-dimensional log Mel-filterbank features, together with an appended ones token at the first timestep, and generates corresponding representations. The first timestep vector of the encoded representations servers as the final acoustic word embedding. This is labeled as $v$ in Fig.~\ref{fig:model}. 

\section{Experiments}
\label{sec:experiment}

\subsection{Training datasets}
\label{ssec:training datasets}
We use the \textit{train-clean-100} and \textit{train-clean-360} split of the publicly available LibriSpeech dataset~\cite{panayotov2015librispeech} for training. The force-alignment transcriptions generated by Montreal Forced Aligner~\cite{mcauliffe2017montreal} are used to get the word boundaries and pairs. The duration of word segments ranges from 0.5 to 2 seconds. During training, the model takes a word pair which is the different acoustic instances of the same word. The true labels of the words are not used. For a word pair $(X, Y)$, we also include $(Y, X)$ in the training. 

\begin{table}[t]
\centering
\centering
\small
\caption{Summary for model architecture and dataset used for CTE Small and Base models.}
\resizebox{\columnwidth}{!}{
\begin{tabular}{lccc}
\toprule

 &   & \textbf{Small} & \textbf{Base} \\ \midrule

Input Feature & & 80-dim FBANK & 80-dim FBANK  \\ 
\midrule

\multirow{5}{*}{Transformer} & layer & 6 & 12 \\
        & embedding dim. & 256 & 512 \\
        & inner FFN dim. & 1024 & 2048  \\
        & attention heads  & 4 & 8 \\
        & average K & 4 & 8 \\
\midrule
Training Set & LibriSpeech & 100 hr & 460 hr \\
\midrule
No. of Word Pairs &  & 121k & 321k  \\
\midrule
Training Steps & & 50k & 60k \\
\bottomrule
\end{tabular}%
\label{tab:model}
\vspace{-0.4cm}
}
\end{table} 

\subsection{Models and training details}
\label{ssec:implementation and training details}
All the models are trained with fairseq toolkit~\cite{ott2019fairseq}. We experiment with two model configurations that share the same encoder architecture but differ in model size. 
Table~\ref{tab:model} summarises all the model parameters for the \textit{small} and \textit{base} models, along with the respective data used. The \textit{small} model consists of 6 Transformer encoder layers, model embedding dimension 256, inner feed-forward network dimension 1024 and 4 attention heads, while the \textit{base} model consists of 12 Transformer encoder layers, model embedding dimension 512, inner feed-forward network dimension 2048 and 8 attention heads. The number of blocks that are used to construct the self-supervised learning targets is 4 and 8 respectively for the CTE \textit{small} and the \textit{base} model. We use 80-dimensional log mel filter bank with a frame length of 25 ms and an overlap of 10 ms as the input features. The models are optimized with Adam~\cite{kingma2014adam}. For the teacher parameterisation, the target decay rate, $\tau$, is set to 0.999. 

\subsection{Evaluation tasks}
\label{ssec:eval datasets}
To assess the quality of the vector representations learned by CTE models, we first conduct a series of analyses on the in-domain LibriSpeech dastaset. We use the CTE learned from the training datasets to encode audio segments in the test sets, which are sourced from LibriSpeech dev-clean and test-clean. It is worth noting that the audio segments used in both training and testing come from in-domain datasets but are mutually exclusive and from different speakers. 

Next, we measure the intrinsic quality of the word embeddings without being tied to a particular downstream task. The same-different task~\cite{carlin2011rapid}, which involves determining whether a given pair of acoustic segments, each representing a true word, are of the same or different word types, is designed for this purpose. The evaluation metric used is the average precision (AP), which is obtained from the area under ROC curve. We conduct the same-different experiments on two different datasets: a cross-domain English dataset and a cross-lingual Xitsonga dataset. For the English dataset, we use speech from Buckeye corpus of conversational English~\cite{pitt2005buckeye}, which contains 6 hours of speech for each of the train, dev and test splits respectively. As for Xitsonga, we use a 2.5-hour portion from NCHLT corpus~\cite{de2014smartphone}. For adapting to the different domain and language, we fine-tune the CTE models on the word pairs obtained from respective datasets. The word pairs are obtained by either ground truth transcriptions or the unsupervised term discovery (UTD) system~\cite{jansen2011efficient}, in which the former sets an upper bound and the latter simulates a low-resource setting. We fine-tune the CTE models for a total of 5k steps in each experiment.

Through these experiments, we evaluate both the generalizing ability of CTE models to a new domain, as well as the transferability of CTE representations to a new unseen language. For a fair comparison, the train, validation and test splits follow the common experimental setting in~\cite{kamper2019truly}. The AP score is reported for all the experiments.

\section{Results and Analysis}
\label{sec:results}

\subsection{In-domain analysis}
\label{ssec:analysis}

\begin{figure}[t]
\begin{minipage}[t]{1.0\linewidth}
  \centering
  \centerline{\includegraphics[width=7cm]{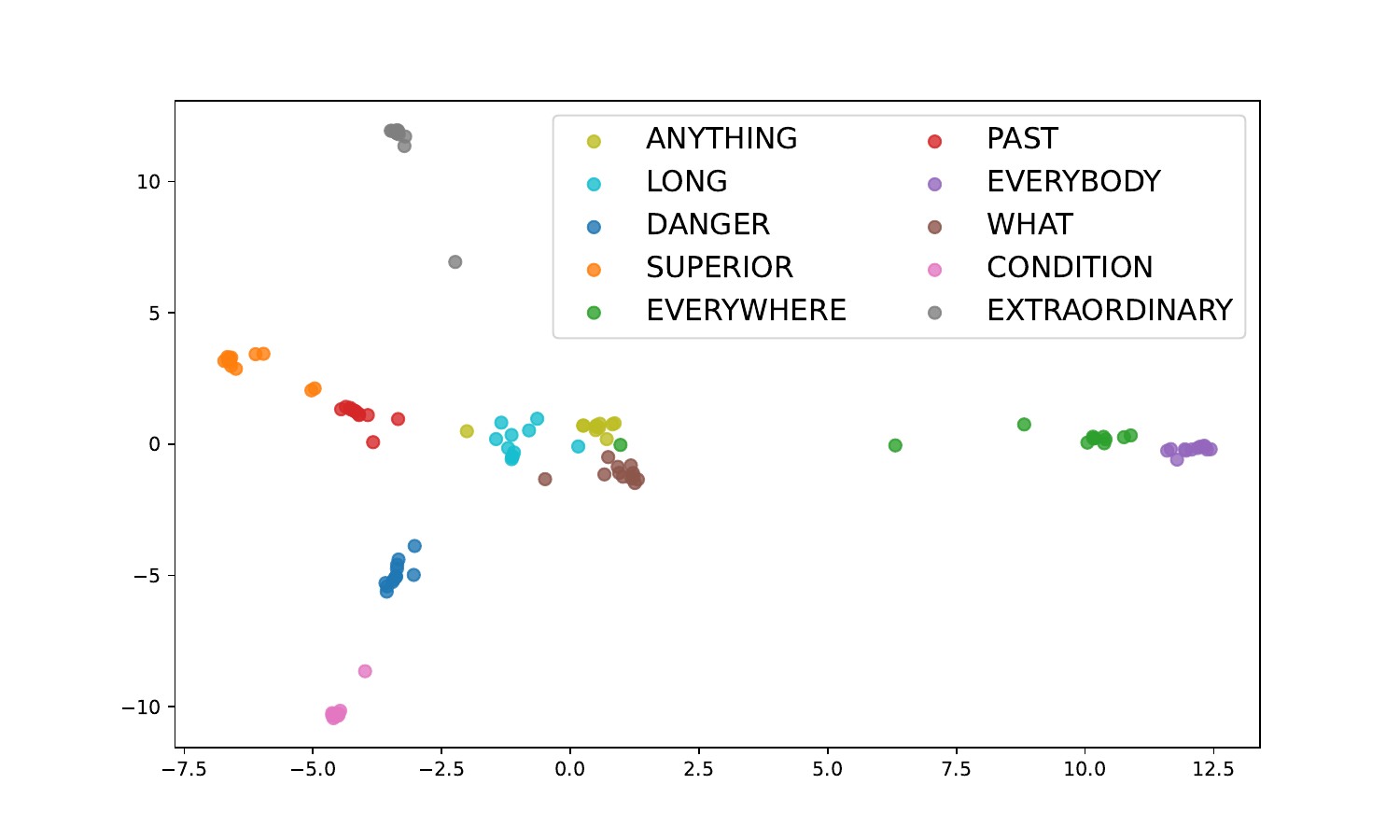}}
  \vspace{-0.2cm}
  \caption{Word embeddings for 10 randomly selected words.}
  \label{fig:random_word}
\end{minipage}
\vspace{-0.6cm}
\end{figure}

Our CTE model aims to map the variable-length speech segments into a fixed-dimensional representation and puts the same word with different acoustic variations close in the underlying space. To demonstrate this, we visualise the word embeddings extracted directly from CTE \textit{base} model. Fig.~\ref{fig:random_word} shows the embeddings extracted from CTE \textit{base} model for 10 randomly selected words from different speakers in the test set described in Section~\ref{ssec:eval datasets}. These embeddings are reduced to two dimensions using principal component analysis (PCA). From the plot, we can see that most of the embeddings of the same words are clustered in close proximity to each other, showing that we have obtained discriminative word embeddings for the unseen speakers. This means our CTE models have learned to disentangle unnecessary speaker information and extracted speaker-independent word-level information. 

\begin{figure}[b]
\vspace{-0.6cm}
\begin{minipage}[b]{1.0\linewidth}
  \centering
  \centerline{\includegraphics[width=7cm]{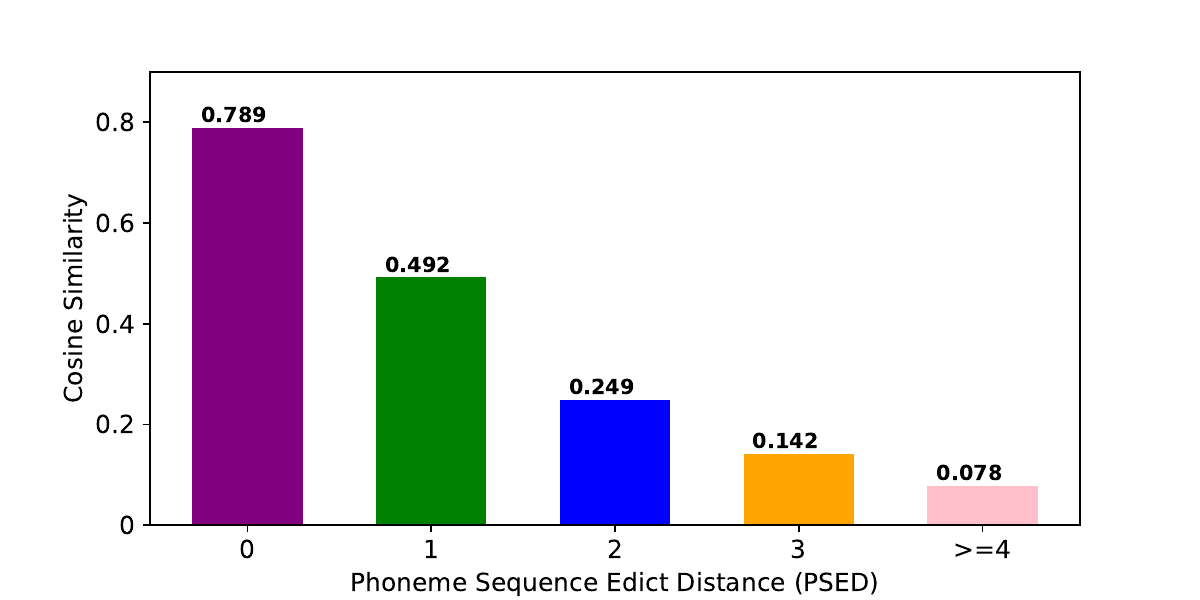}}
  \caption{The average cosine similarity between embeddings for different phoneme sequence edit distance (PSED).}
  \label{fig:psed}
\end{minipage}
\vspace{-0.6cm}
\end{figure}

Next, we analyse if the embeddings obtained can effectively distinguish words with similar phonetic structures. We compute the average cosine similarity for each pair of acoustic segments against the phoneme sequence edit distance (PSED), shown in Fig.~\ref{fig:psed}. From the plot, it is obvious that word pairs with larger PSED have smaller cosine similarities. Notably, the cosine similarity is largest for PSED = 0, and there is a clear drop from 0.789 to 0.492 as PSED increases from 0 to 1. This clear distinction is useful for applications that require high recall. As PSED increases to 4 or more, the average cosine similarity drops to 0.078 only. The gradual decrease in cosine similarity with increasing PSED also indicates that the word embeddings are able to describe the sequential phoneme structure despite being trained with only word-level loss.

\subsection{Cross-domain analysis}
\label{ssec:cross-domain}

\begin{table}[b]
\centering
\vspace{-0.3cm}
\centering
{\footnotesize
\caption{Average precision on Buckeye English validation set and test set for different models using ground truth word pairs. As we take the statistics from the papers directly, some average precision of the test sets are not reported in their papers.}
\begin{tabular}{l|c|c}
\toprule

\textbf{Model} & \textbf{Validation} & \textbf{Test}  \\
\midrule
CTE small                  & 67.0  & 71.7 \\
CTE base                   & \textbf{73.3}  & \textbf{76.3}   \\
\midrule
Downsampling~\cite{kamper2019truly}             & 24.5  & 21.7 \\
DTW alignment~\cite{kamper2019truly}            & 36.8  & 35.9 \\
ENCDEC-CAE~\cite{kamper2019truly}               & 51.1  & - \\
MCVAE~\cite{peng2020correspondence}             & 58.8  & -  \\
HuBERT (EN) mean~\cite{sanabria2022analyzing}   & 67.5  & 67.8 \\
HuBERT (EN) subsample~\cite{sanabria2022analyzing}  & 65.3 & 64.8 \\
\bottomrule 

\end{tabular}%
\label{tab:same-diff-gt}
}
\end{table} 

\subsubsection{Ground-truth upper bounds}
In this section, we evaluate the generalizability of the proposed CTE model across a different domain. Table~\ref{tab:same-diff-gt} presents the average precision (AP) score on Buckeye English validation and test sets for different models, with training word pairs obtained from ground-truth transcriptions. As a strong baseline, we include HuBERT~\cite{hsu2021hubert}, a self-supervised learning model that is pre-trained on the 60k hour split from the Libri-light dataset~\cite{kahn2020libri}. The word embeddings are obtained by either mean pooling or subsampling the representations from HuBERT Large~\cite{sanabria2022analyzing}. Except for HuBERT, other baseline models are trained on the same word pairs from the Buckeye datasets.

As reported in Table~\ref{tab:same-diff-gt}, the strongest baseline models are HuBERT and MCVAE~\cite{peng2020correspondence}. Specifically, HuBERT with mean pooling has achieved an AP score of 67.5\% and 67.8\% on validation and test sets, while MCVAE has achieved an AP score of 58.8\% on the validation set. 
In comparison, CTE models show substantial improvements over all other embedding approaches: CTE \textit{small} achieves 67.0\% and 71.7\% and CTE \textit{base} achieves 72.3\% and 75.5\% on the validation and test sets. This shows that the embeddings learned capture acoustic information that is general and transferable to other domains. 
Moreover, the fact that our CTE models outperform HuBERT, which is pre-trained on a much larger dataset, also implies that our word-level training strategy is more robust to variations and effective in capturing word-level information than the frame-level training strategy that is adopted in HuBERT. 

\subsubsection{Unsupervised setting}
\label{sssec:unsupervised_buckeye}
Table~\ref{tab:same-diff-utd} presents the results for training using pairs discovered by the UTD system~\cite{jansen2011efficient}, which simulates the low-resource setting. In this setting, the training pairs are discovered by a UTD system, while a small set of ground truth pairs are used as the validation data. The training pairs discovered by the UTD system inevitably suffer from erroneous matches. Therefore, instead of using complex models to learn from potentially inaccurate training pairs, our models can leverage prior knowledge to mitigate the effects of inaccuracy. The experimental results show that the CTE \textit{base} model exhibits promising performance in the low-resource setting, achieving an AP score of 43.2\% on validation set and 44.1\% on test set.



\begin{table}[t]
\centering
\centering
{\footnotesize
\caption{Average precision on the cross-domain Buckeye English validation and test datasets, as well as the cross-lingual Xitsonga test set for different models. The training pairs for both datasets are discovered using the unsupervised term discovery system (UTD).}
\begin{tabular}{l|c|c|c}
\toprule
\textbf{Model} & \multicolumn{2}{c|}{\textbf{English}} & \textbf{Xitsonga} \\
& \textbf{Valid} & \textbf{Test} & \textbf{Test} \\
\midrule
CTE small                  & 32.9  & 32.8 & 31.1 \\
CTE base                   & \textbf{43.2}  & \textbf{44.1} & \textbf{46.7} \\
\midrule
Downsampling~\cite{kamper2019truly}                            & 24.5  & 21.7 & 13.6 \\
DTW alignment~\cite{kamper2019truly}                           & 36.8  & 35.9 & 28.1 \\
CAE-RNN~\cite{van2021comparison}                              & -     & 36.8 & 41.8 \\  
ENCDEC-CAE~\cite{kamper2019truly}                             & 31.7  & 32.2 & 32.0  \\
MCVAE~\cite{peng2020correspondence}                           & 37.6  & 39.5 & 44.4 \\
HuBERT (EN) mean~\cite{sanabria2022analyzing}                 & -     & -    & 39.0 \\
HuBERT (EN) subsample~\cite{sanabria2022analyzing}            & -     & -    & 46.0 \\
\bottomrule

\end{tabular}%
\label{tab:same-diff-utd}
}
\end{table}

\subsection{Cross-lingual analysis}
The above experiments have shown promising results in English even when in-domain data is scarce. However, there are many low-resource languages where a large corpus might not be available. Therefore, our objective here is to investigate the transferability of CTE models trained in English when employed in a different language. In Table~\ref{tab:same-diff-utd}, we show the performance of word embeddings obtained from the CTE models when applied to Xitsonga, an unseen low-resource language. Among all the models, the CTE~\textit{base} model achieves an AP score of 46.7\%, which is comparable with HuBERT with a subsampling pooling strategy. However, note that HuBERT with subsampling pooing strategy produces a word embedding with a much larger dimensionality (10240 dimensions), which is less efficient for applications with limited memory.

Interestingly, the CTE models, when evaluated on Xitsonga, outperform itself when evaluated on Buckeye English. This trend can be attributed to the higher pair-wise matching precision in Xitsonga than in Buckeye English. Consequently, this finding suggests that the CTE model is robust to changes in language while still being sensitive to the quality of training pairs, despite the beneficial effects of pre-training in mitigating the impact of inaccurate training pairs as shown in Section~\ref{sssec:unsupervised_buckeye}.





\section{Conclusion}
\label{sec:conclusion}
In this paper, we propose a novel
self-supervised framework, CTE, to learn robust acoustic word embeddings from the large-scale unlabeled speech corpus and achieves state-of-the-art performances.
Our experiments have shown that the word embeddings extracted from the proposed CTE model exhibits different levels of robustness, particularly in their robustness towards the change of speakers, domain and language, which positions it as a competitive model in learning acoustic word embeddings. 
Future work includes scaling up the model size and training data to increase the model capacity, improving the quality of word pairs obtained from the UTD system and applying the CTE models to more diverse downstream tasks.

\section{Acknowledgements}
This work is supported by
1) Huawei Noah's Ark Lab;
2) National Natural Science Foundation of China (Grant No. 62271432);
3) Agency for Science, Technology and Research (A*STAR) under its AME Programmatic Funding Scheme (Project No. A18A2b0046)

\bibliographystyle{IEEEtran}
\bibliography{mybib}

\end{document}